\documentclass[preprint]{elsarticle}
\usepackage{graphicx}
\graphicspath{{figures/}}
\usepackage{amsmath}
\bibpunct[, ]{[}{]}{,}{a}{,}{,}
\ifnum\lefthyphenmin<2\lefthyphenmin=2\fi
\ifnum\righthyphenmin<2\righthyphenmin=2\fi

\begin{document}

\begin{frontmatter}

\title{Nanoepitaxy on quasicrystal surfaces}
\author[eth,bu]{M.~Erbudak\corref{cor1}}
\ead{erbudak@phys.ethz.ch}

\author[bu,fge]{M.~Mungan}
\ead{mmungan@boun.edu.tr}

\author[eth]{S.~Burkardt}
\ead{svenbu@phys.ethz.ch}

\cortext[cor1]{Corresponding author}
\address[eth]{ETH Zurich, 8093 Zurich, Switzerland}
\address[bu]{Bo\u gazi\c ci University, Bebek, 34342 Istanbul, Turkey,}
\address[fge]{The Feza G\"ursey Institute, P.O. Box 6, \c Cengelk\"oy, 34680 Istanbul, Turkey}

\begin{abstract}
In film growth on quasicrystalline surfaces, the epitaxy-imposed ordering cannot compete with the stable bulk phases of thick films due to absence of translational order in the structure of the substrate. Energetically, this renders the formation of crystalline domains in the native structure of the film material more favorable while their global orientation is prescribed by the quasicrystalline order. We present experimental results on the dissociative chemisorption of oxygen on the decagonal surface of Al$_{70}$Co$_{15}$Ni$_{15}$ as well as  molecular-dynamics simulations of the diffusion of adatoms on the surface of the partially covered substrate.
\end{abstract}


\end{frontmatter}

\section {Introduction}
In heteroepitaxial growth, the properties of the film strongly depend on the amount of structural mismatch with the substrate as well as on the film thickness. The latter adversely influences the elastic energy of the strained film giving rise to a three-dimensional growth regime and the formation of islands with limited sizes. This phenomenon can be utilized to fabricate nanometer-size domains by depositing material on a substrate with a large lattice mismatch of typically in excess of 5 \% \cite{pehlke}. An extreme case is encountered by growing crystalline material on a quasicrystalline substrate for which no translational periodicity exists \cite{sharma}. We will outline below our experimental and computational efforts to elucidate the processes encountered in such an extreme case of lattice-mismatch in heteroepitaxy.

The decagonal quasicrystal Al$_{70}$Co$_{15}$Ni$_{15}$ ($d$-AlCoNi) has a structure which is quasiperiodic in two dimensions, periodic in the third dimension, and displays decagonal diffraction symmetry. Investigations of its bulk structure have revealed a columnar prismatic morphology with the column axis parallel to the periodic direction, the tenfold-symmetry axis \cite{refik2,steurer2}. The decagonal surface is perpendicular to the tenfold-symmetry axis and is very suitable as a substrate for epitaxial growth due to its flatness \cite {kishida,yuhara}. 

We have previously reported that Al adsorption on the tenfold-symmetry surface results in 3-nm large Al(111) domains for coverages in excess of a few monolayers (ML). The azimuthal orientation of the domains is governed exclusively by the aperiodic symmetry of the substrate \cite{Thomas}. We also have studied numerically the effect of the relative strength of mutual interactions of adsorbed atoms (adatoms) with respect to their interactions with the substrate atoms \cite{bilki,mungan} on the structure of the growing adsorbate layers  and how the structure of the adsorbate layer (adlayer) depends on the rate at which adatoms thermally equilibrate with the substrate \cite{sven}. Deposition is a non-equilibrium process and the ratio of the thermal relaxation to the deposition rate of the adatoms has influence on the resulting morphology of the adsorbate \cite{barabasi}. We found in particular that the thermal relaxation rate and relative strength of the interaction of adsorbate atoms with each other and the substrate was able to change  the surface growth mode  from cluster to layer-by-layer type. We attributed this to (i) the extent to which arriving adatoms can diffuse on the surface before being thermalized and (ii) how strongly already present adatoms can steer this diffusion process. Our purpose in this report is to further elucidate this behavior by focusing on the diffusion of deposited adatoms on the substrate-adsorbate surface. We also present experimental results on the initial stages of the high-temperature oxidation of the tenfold-symmetry surface of $d$-AlCoNi resulting in the formation of hexagonal domains of dissociated oxygen. The thickness of the oxygen adlayer as well as its properties are determined by the conditions at which oxygen interacts with the surface. Hence, this is a self-regulating system for which the physics and chemistry at the interface determine the final surface constellation.

\section {Experimental}

In this report we have used a $d$-AlCoNi quasicrystal with dimensions 5\hspace{1pt}$\times$\hspace{1pt}3\hspace{1pt}$\times$\hspace{1pt}1 mm$^3$ which was mounted on a resistive heater, while the sample temperature was measured with a chromel-alumel thermocouple (K-type) pressed onto the sample surface \cite{kortan2}. The quasicrystal surface was cleaned in ultrahigh vacuum by cycles of sputtering with Ar$^+$ ions (1.5 keV, 5\hspace{1pt}$\times$\hspace{1pt}10$^{-7}$ A/mm$^2$) at 670 K and heat treatment at 900 K for 30 minutes. The sample preparation was monitored by the quality of the low-energy electron diffraction (LEED) pattern and by inspecting the scans of Auger electron spectroscopy (AES). A three-grid, back-view, display-type LEED system, operating at low microampere primary currents, had a total opening angle of $102\pm2^{\circ}$ \cite{ociepa}. Thus, a momentum transfer of $2.83\pm 0.04$ \AA$^{-1}$ could be detected for 50 eV electrons at normal incidence.

The diffraction patterns were recorded by a 16-bit charge-coupled device camera after the sample was cooled down to room temperature. The position and the size of the Bragg spots in LEED observations are used to extract real-space information about the atomic structure, the size, and the orientation of the surface textures. The AES spectra were recorded using a cylindrical mirror analyzer operating with a constant relative resolution of $\Delta E/E = 0.8 $\%. Experimental details have previously been reported \cite{review,Thomas,Longchamp1,sven}.

\begin{figure}[!t]
\includegraphics[width=0.8\columnwidth]{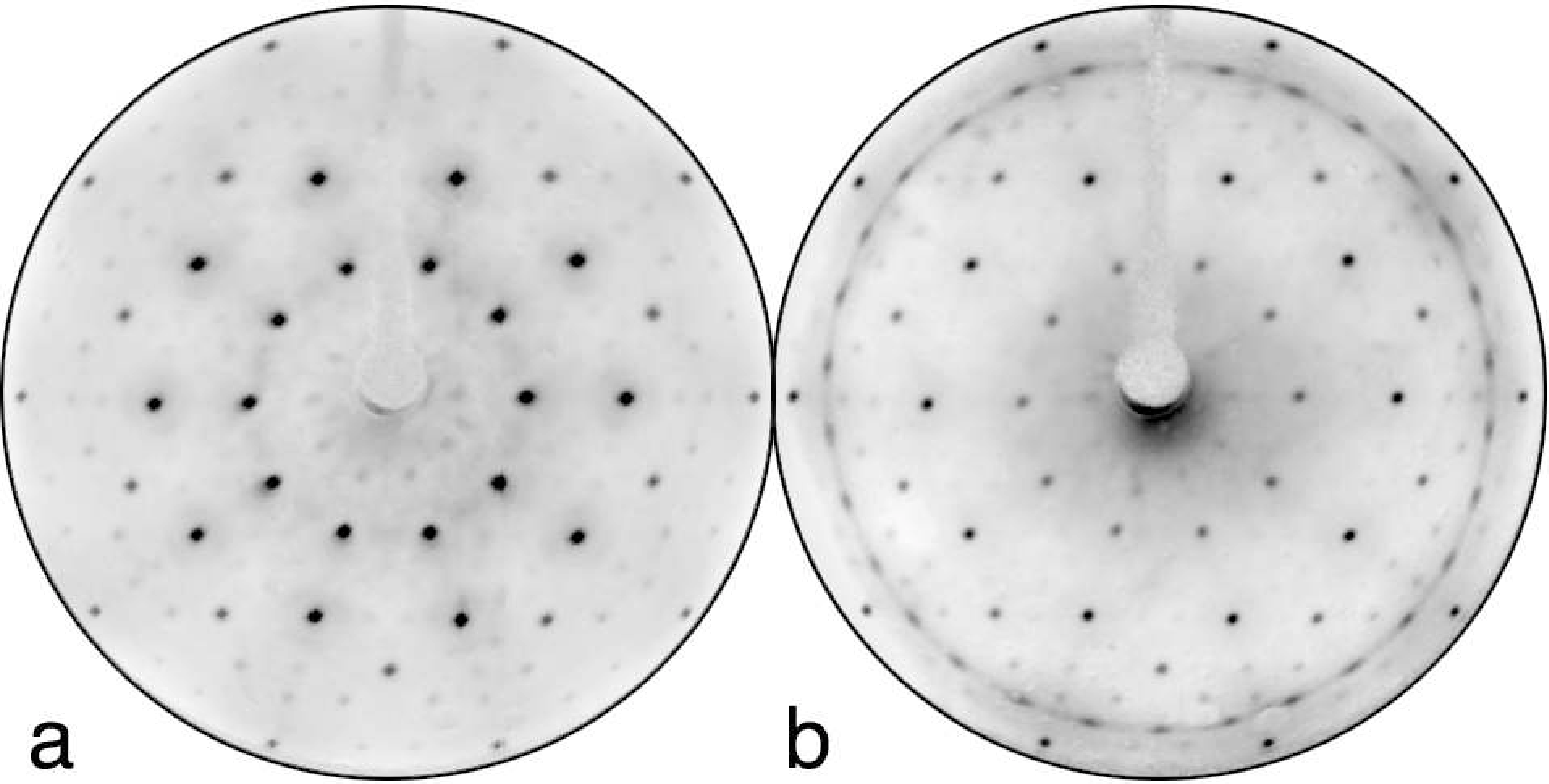} 
\caption{LEED patterns at a primary-electron energy of 50 eV obtained from (a) the clean and (b) the exposed tenfold-symmetry surface of $d$-AlCoNi to 100 Langmuirs of oxygen at 870 K. The patterns are shown after normalization by the response function of the display and recording system in order to eliminate spurious contributions to the image.}
\label{fig:leed}
\end{figure}

\section {Oxygen adsorption at high temperatures}

Fig.~\ref{fig:leed}a shows a LEED pattern from the clean tenfold-symmetry surface of $d$-AlCoNi at 50 eV. The pattern displays tenfold symmetry with spot profiles having a total width of about 0.5$^{\circ}$ corresponding to an average terrace size of at least 150 \AA. This value is comparable with the coherence length of the primary electrons used here and, therefore, is the upper limit of the domain size that can be determined in our apparatus \cite{ociepa}.

The LEED pattern depicted in Fig.~\ref{fig:leed}b is obtained from the surface after exposure to oxygen partial pressure of 1\hspace{1pt}$\times$\hspace{1pt}10$^{-8}$ mbar at 870 K for 1000 s. We discern thirty new diffraction features placed on a polar circle of about $40.5\pm0.5^{\circ}$. In accordance with the local symmetry of the quasicrystalline surface, these spots represent a sixfold-symmetric pattern, repeated five times in equal azimuthal increments of $2\pi/5$. Each sixfold-symmetric pattern is due to a hexagonal mesh with an interatomic distance of $3.08\pm0.03$ \AA, as extracted from LEED. We note that his value is $10$\% larger than the interatomic O-O distance in different phases of Al$_2$O$_3$ \cite{jaeger}. A lattice expansion of a similar amount has also been encountered in oxide films grown on Al$_{70}$Pd$_{20}$Mn$_{10}$ \cite{Longchamp1}. It is important to note that the intensity of the new diffraction features do not show any polar smearing compared to those from the quasicrystal, while they are extended azimuthally by $2-3^{\circ}$. We may interpret the polar precision as an unexpectedly large domain size of about 150 \AA, while the azimuthal spread arises as a consequence of the relaxation of the crystalline adlayer compared to the quasicrystalline substrate. The latter facilitates the growth of the adlayer in this extraordinarily large size.
We note that the diffraction spots characteristic of the clean quasicrystalline surface are still observable indicating that the surface film is rather thin (few \AA), even thinner than 5 \AA, the thickness of oxide films grown on the fivefold-symmetry surface of Al$_{70}$Pd$_{20}$Mn$_{10}$ \cite{Longchamp1}. The observation of these spots at the same place further indicates that the formation of the surface layer preserves the quasicrystalline order at the interface. 
The azimuthal alignment of the spots with respect to the substrate is shifted by $6^{\circ}$ compared to that observed for Al films deposited onto the same decagonal surface \cite{Thomas}.

\begin{figure}[!t]
\includegraphics[width=0.8\columnwidth]{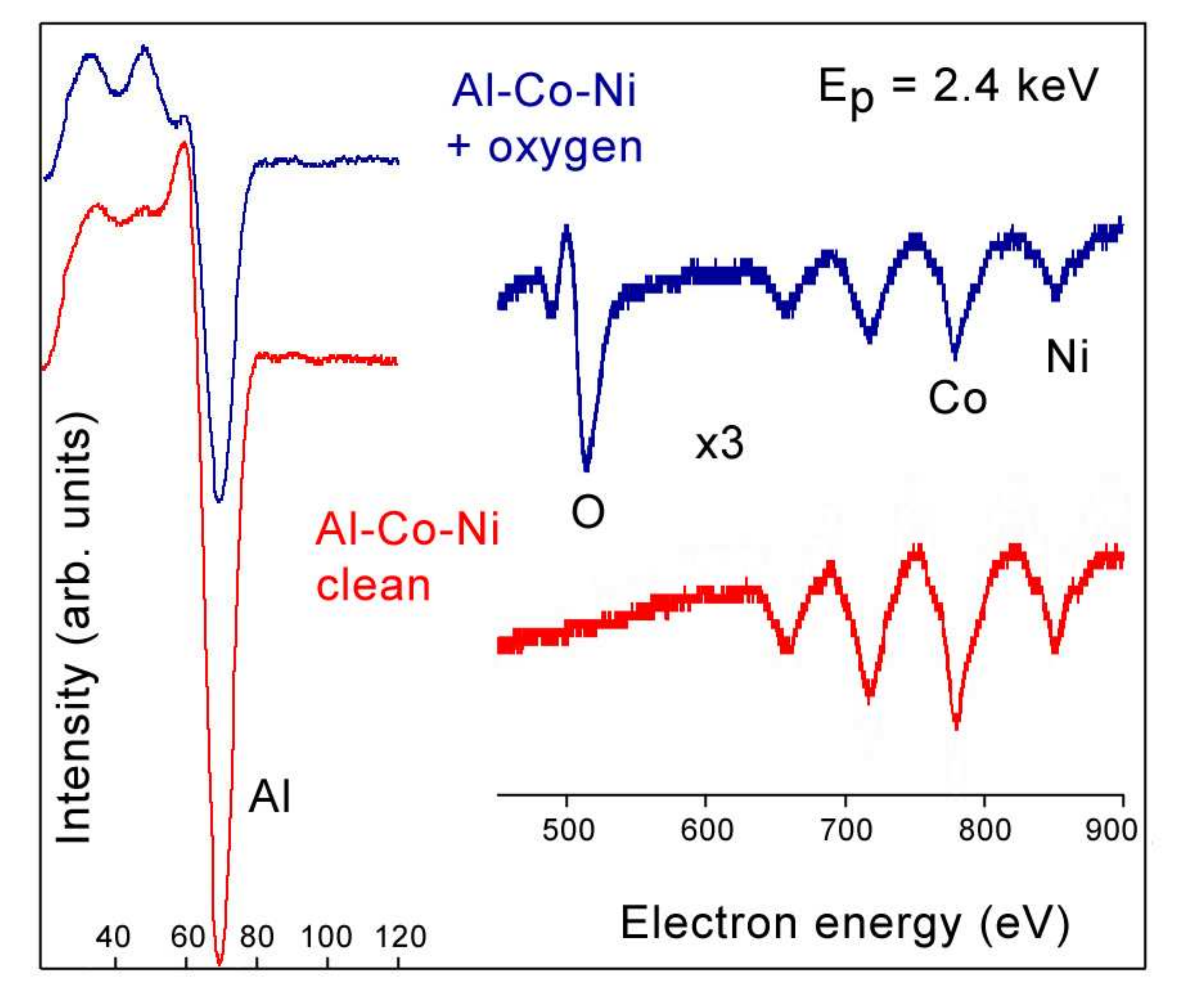} 
\caption{Auger spectra obtained from the clean (red, bottom) and oxygen-adsorbed (blue, top) tenfold-symmetry surface of $d$-AlCoNi in the energy region of the Al as well as O, Co, and Ni transitions. Spectra for the high-energy region are enlarged by a factor of 3.}
\label{fig:aes}
\end{figure}

Fig.~\ref{fig:aes} shows Auger spectra near the Al $2pVV$, O $1sVV$, Co and Ni $2p3d3d$ transitions for the clean (bottom) and oxygen-exposed (top) surface at 2.4 keV excitation energy. The spectrum obtained from the clean surface is dominated by the Al transition at around 70 eV similar to that obtained from clean Al. For an oxidized Al sample this peak is found shifted to lower energies by almost 10 eV due to a strong electron transfer from Al to O \cite{alox}. In the present case, however, no energy shift for the Al signal is observed after oxygen adsorption, but some reduction of the intensity. This observation indicates that there is no appreciable electron transfer from the metal site to oxygen, i.e., oxygen is in the chemisorbed state. The tiny shoulder at around 60 eV in the top spectrum may signal the presence of a minute amount of aluminum oxide and hence the early stage of oxidation. Around 500 eV the spectrum from the oxygen-exposed surface shows the oxygen $1sVV$ signal. Both surfaces display $2p3d3d$ transitions for Co and Ni at higher kinetic energies. Except for a slight intensity reduction, no chemical effects can be discerned in this energy region for the transition-metal signals indicating that oxygen binding to Co or Ni is negligible.

Taking into account the features of the LEED pattern and the results of AES, the adsorbed oxygen may be just one single ML thick. It is interesting to note that while a ML of noble gas forms a pseudomorphic layer \cite{xenon}, oxygen adsorption distinctly forms a hexagonal lattice on the quasicrystalline surface.

\section {Simulation of adatom diffusion on the adsorbate-substrate surface}

We perform molecular-dynamics simulations of how deposited adatoms diffuse on the adsorbate-substrate surface  in order to understand the morphology of the growing adlayer. We use a 6\,000-atoms substrate model of the decagonal Al$_{70}$Co$_{15}$Ni$_{15}$ quasicrystal with dimensions  $150\times150\times 2.05$ \AA $^3$   \cite{sven,sofia10}. The large lateral size of the unit cell reduces artefacts generated by boundary conditions. For the dimensions given above, 1 ML of adatoms corresponds to about 3100 particles. In the simulations we inject adatoms in batches of 180 particles onto the substrate. The quasicrystalline bilayer is kept rigid, but adatoms are allowed to interact with each other and with the substrate atoms. The relative strength of the adatom-adatom and adatom-substrate interactions is controlled by a dimensionless parameter $\eta$. Experimentally, it is found that the adsorption of Al on AlCoNi yields qualitatively similar results as the adsorption of nobel gases such as Xe on the same type of substrates \cite{Thomas,xenon}. These findings imply:  (i) given that Al and Xe have very different 
chemical binding properties, the mechanism of domain formation 
does not depend strongly on the details of the interaction, which motivates our use of a simple 
Lennard-Jones interaction potential (ii), the interaction between adsorbate Al atoms and Al in the substrate might be different from the Al interactions in crystalline Al and thus the  
strengths of the interaction of Al adatoms with atoms in the adsorbate and substrate are not necessarily 
equal. 

In real experiments, hot impinging adatoms loose their excess kinetic energy to the substrate and eventually reach thermal equilibrium. 
In our simulations, we account for this process by introducing damping and treating the damping rate $\gamma$ as a parameter by which we can change the thermal relaxation times. Numbers given below are in units in which lengths are measured in angstrom and the mass of the adatoms is set to one.  

Numerical details of the implementation are as follows. We let Adatom particles $i$ interact with each 
other and the atoms $\alpha$ of the rigid quasicrystalline substrate via a Lennard Jones potential,  
\begin{equation}
V(r) = 4 \epsilon \left [ \left (\frac{\sigma}{r} \right )^{12} 
- \left ( \frac{\sigma}{r} \right )^6 \right ],  
\end{equation}
with parameter values of $\sigma = 2.55$ (in accordance with the nearest neighbor distance of Al atoms in the fcc structure) and $\epsilon = 0.25$, which establishes an energy scale (eV). The equations of motion of 
the adatoms are given by.
\begin{equation}
\ddot{\bf r}_i = - \sum_{\alpha} {\bf \nabla} V(|{\bf r_i - r_\alpha}|) 
                   - \eta \sum_{j < i} {\bf \nabla} V(|{\bf r_j - r_i}|) + {\bf f}_i(T), 
\label{eom}
\end{equation}
where the parameter $\eta$ is the relative strength of the adatom interactions with respect to the substrate 
and ${\bf f}_i(T)$ is the damping term for the thermal relaxation
\begin{equation}
{\bf f}_i(T) = - \gamma \left ( \frac{1}{2} (\dot{\bf r}_i)^2 - \frac{3}{2}T \right ) {\bf \hat{\dot{r}}}_i, 
\end{equation}
with $\gamma$ being the thermal damping rate, $T$ the temperature measured in units of $\epsilon$ and 
${\bf \hat{\dot{r}}}$ being the velocity unit vector of the particle \cite{Gilmore}. We do not distinguish 
between the different atomic types in the substrate. 
The equations of motions are integrated using an adaptive step size fourth order Runge-Kutta scheme, \cite{NumRec}. Given the relaxational nature of the particle dynamics, the choice of an adaptive steps size controller turned out to be an efficient way of maintaining numerical precision at a minimum number of integration steps. 

\begin{figure}[!t]
\includegraphics[width=0.8\columnwidth]{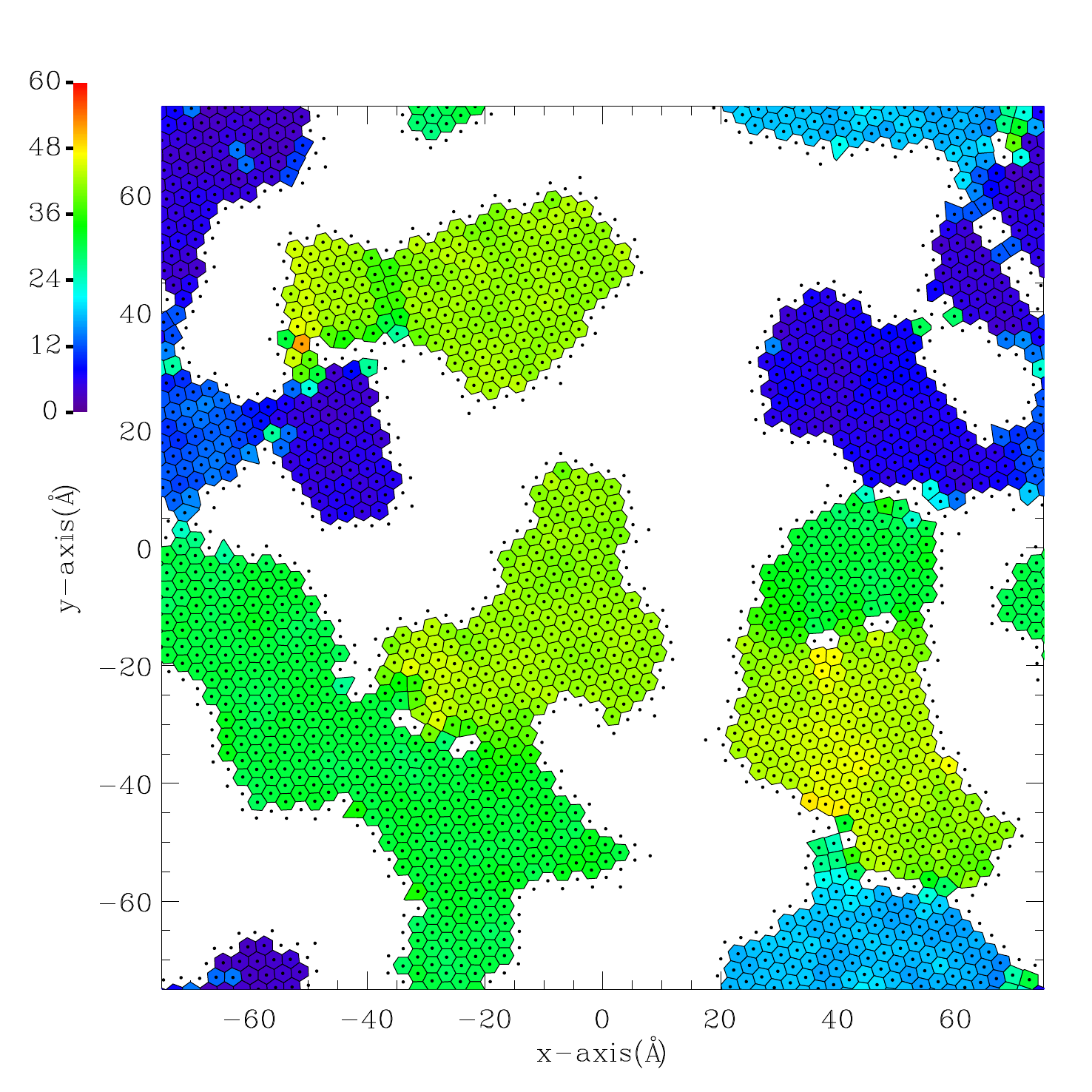} 
\caption{Voronoi tessalation of the third adlayer, as counted from the quasicrystal interface  at 3.5 ML coverage, low damping rate $\gamma = 0.0625$, and strong adatom interactions $\eta = 2$. Voronoi cells associated with atoms exhibiting close to perfect local sixfold coordination have been color-coded according to the angle that the corresponding hexagons make with the x-axis. Notice the formation of nano-scale well-ordered domains with fixed orientations of $6^\circ, 18^\circ, 30^\circ$, and $42^\circ$. The fixed orientational angles are a direct consequence of being in registry with the tenfold symmetry of quasicrystalline substrate. Another possible domain orientation of 56$^{\circ}$ is absent.}
\label{fig:grains}
\end{figure}

\begin{figure}[!t]
\includegraphics[width=1.0\columnwidth]{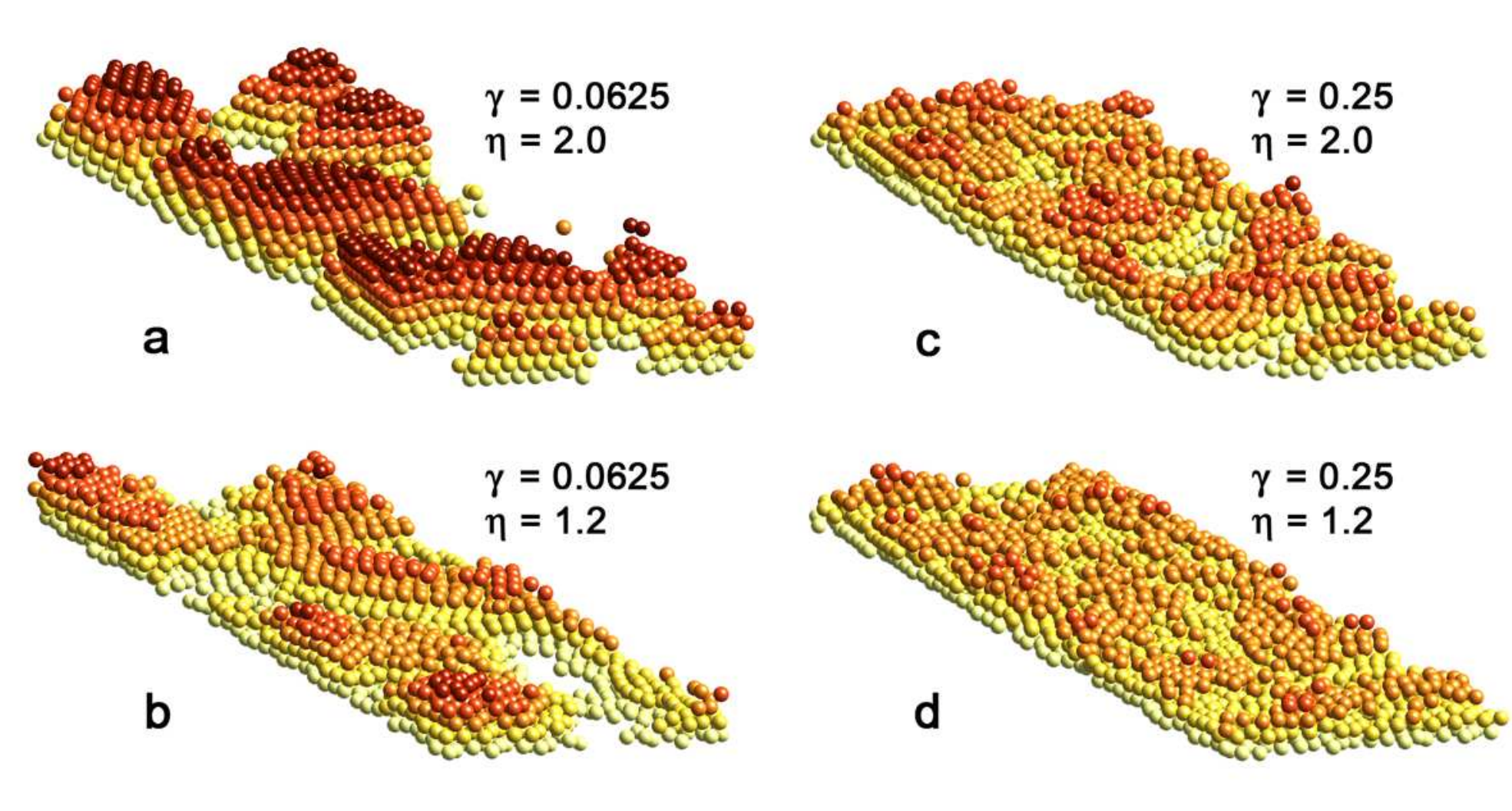} 
\caption{Structure of the adsorbate surface at 2.5 ML coverage at a damping rate of $\gamma = 0.25$ at different strength of the adatoms interactions relative to the substrate. The top bottom panels show the resulting adlayer morphology for $\eta = 2$ and 1.2, respectively.  Notice the more locally ordered structure for stronger adatom interactions ($\eta = 2$). Also, for $\eta = 2$ the adlayer is more corrugated.}
\label{fig:section}
\end{figure}

We have already shown that by varying $\gamma$ and $\eta$, different surface growth morphologies including layer-by-layer and cluster-type growth can be observed \cite{mungan}. We attributed this finding to the fact that a higher damping rate, which corresponds to a short thermal relaxation time, causes the particles to hit the substrate and stick, very similar to rain drops wetting a surface, while the effect of varying $\eta$ determines how strongly already deposited adatoms can steer impinging adatoms during their diffusive motion \cite{sven}. We were able to explain how deposition at  high $\gamma$  along with low  $\eta$ values, {\it e.g.}, $\gamma$ = 0.25 and $\eta = 1$, gives rise to layer-by-layer type of growth, while keeping $\gamma$ constant and increasing $\eta$ will increase the probability of arriving adatoms to be deposited on already present ones giving rise to a cluster-type growth as observed in our simulations. We also found that well-ordered domains can grow at lower $\eta$ values if the damping rate is reduced, so that impinging adatoms can diffuse longer on the substrate surface and, moreover, with increasing coverage, there will be more accessible favorable binding sites, since they will include both already deposited adatoms as well as substrate sites that have remained exposed \cite{sven}. Fig.~\ref{fig:grains} demonstrate the formation of ordered and orientationally locked local hexagonal domains. We have performed a Voronoi tesselation of the the third adlayer above the interface and color-coded the Voronoi hexagons associated with sixfold-coordinated atoms that show close to perfect local order also in the bond-angles. The color of each hexagonal Voronoi cell is determined by the angle relative to the x-axis and corresponds to a range from $0$ to $60^{\circ}$. Multiple domains with relative orientation of $6^\circ, 18^\circ, 30^\circ$, and $42^\circ$  are readily discerned. The fixed orientation angles of the domains are a result of the registry with the underlying tenfold symmetry of the quasicrystalline substrate. An excerpt of the three-dimensional configuration of the adsorbate film is shown in Fig.~\ref{fig:section}.  
  
Here we aim to further substantiate these observations. Specifically, we analyze the deposition and the subsequent lateral diffusion of an additional set of 180 adatoms (0.06 ML) injected onto the adsorbate surfaces when 0.5, 1.5, and 2.5 ML of adatoms have already been deposited.  We ask how the lateral diffusion of impinging adatoms depends on the already present coverage, the damping rate $\gamma$, and the relative strength of the adatom interactions $\eta$. We track the lateral diffusion of the impinging adatoms over a time scale sufficiently long to reach thermal equilibrium with the substrate and measure the displacement from the initial lateral positions as a function of time.

\begin{figure}[!t]
\includegraphics[width=0.8\columnwidth]{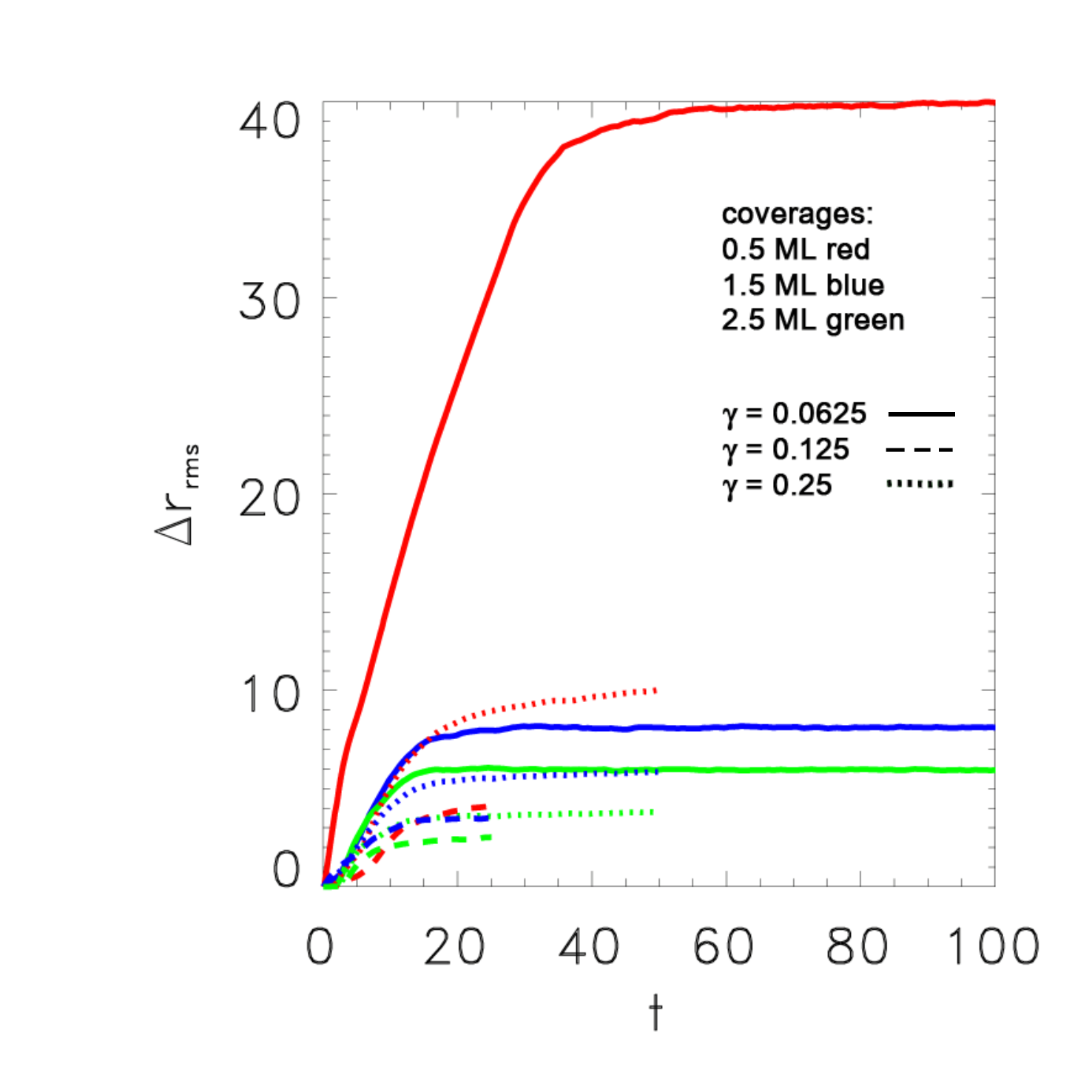} 
\caption{Root-mean-squared lateral diffusion distance against time. Shown are curves from three different damping rates $\gamma$ and and surface coverages: $\gamma = 0.0625$ (solid curves), $\gamma = 0.125$ (dashed curves), and $\gamma = 0.25$ (dotted curves), for coverages of 0.5 (red), 1.5 (blue), and 
2.5 (green) ML. See text for further details.}
\label{fig:diffusion}
\end{figure}

Fig.~\ref{fig:diffusion} shows the root mean squared lateral displacement $\Delta r_{\rm rms}$ averaged over impinging adatom trajectories that stick to the substrate surface as a function of time for $\eta = 2$ and $\gamma$ values of 0.0625, 0.125, and 0.25 depicted as solid, dotted and dashed lines, respectively. The dependence of the diffusion on the coverage of already deposited atoms on the surface is shown, color-coded as follows:
0.5 ML (red), 1.5 ML (blue), and 2.5 ML (green). We clearly see that the maximal diffusion distance decreases sharply with increasing thermal relaxation rate, as one expects. On the other hand, for a low relaxation rate ($\gamma = 0.0625$, solid lines), the diffusion distance drops sharply from 40 \AA\ at a coverage of 0.5 ML to 8 \AA\ at 1.5 ML along with a lesser drop to 6 \AA\ at 2.5 ML, while the dependence of the diffusion distance on the substrate coverage becomes less pronounced for higher damping rates. 

It turns out that for $\eta = 1.2$ almost the same diffusion is obtained at 
1.5 and 2.5 ML and coverages, while the behavior at 0.5 ML coverage yields shorter distances. Thus, in general, it seems that the effect of $\eta$ on the diffusion range is less pronounced or present only at low coverages.  Nevertheless, we know from our previous work that the adsorbate covered surfaces reveal different morphologies. In Fig.~\ref{fig:section} we show $150\times50\times15$ \AA $^3$ sections of the adsorbate  for values of $\gamma = 0.0625$ (left column)  $\gamma = 0.25$ (right column) and $\eta = 2$ (top row) and $\eta = 1.2$ (bottom row) after deposition of 2.5 ML. Although we find comparable diffusion distances for configurations with different $\eta$ (configurations in the same column), the morphology of the corresponding surfaces is different. For both $\gamma = 0.25$ and 0.0625, the surfaces with stronger relative adatom interactions, $\eta = 2$ exhibit enhanced well-ordered hexagonal domains, which are less pronounced for $\eta = 1.2$ (bottom row). 
The adlayer formed under low damping and strong adatom interaction shows 
the largest amount of corrugation along with local hexagonal order (top left). 

\section{Discussion and Conclusions}

The dissociative adsorption of oxygen has already been observed on Al(111) \cite{Flod}, where the formation of a hexagonal oxygen structure is compatible with the substrate symmetry. In the present case, however, the substrate, while containing Al, has a quasicrystalline structure on which the sixfold symmetric oxygen adlayer assembles. While the Al-O interface for Al(111) is stable (nonreactive) up to temperatures of 470 K, Al in the quasicrystalline Al-Co-Ni matrix remains nonreactive at temperatures close to 900 K.

We observe the formation of unusually large oxygen islands on the quasicrystalline substrate. The influence of the latter, beyond providing the correct physical and chemical environment for the formation of the novel oxygen islands, consists of aligning these islands in a quasicrystalline matrix. Each island is quite large and low-energy electrons used in imaging the surface do not show a coherent interaction between the islands.

In the simulations, we have used a quasicrystalline bilayer perpendicular to the periodic direction of the structural model to investigate the lateral diffusion of adatoms as they search for energetically 
favorable locations. We find that the morphology of the developing adlayer can be explained 
in terms of the thermal relaxation times for the impinging adatoms to reach thermal equilibrium 
and the relative strength of the adatom interactions with each other and the substrate. 
Short thermal relaxation times generally lead to a layer by layer growth mechanism, 
in which the randomly incident adatoms hit the substrate and diffuse very little. The diffusion 
range is also found to diminish with the coverage, but this behavior is less pronounced at 
small thermal relaxation times. For coverages of 1.5 ML and more, the relative strength of the 
adatom interaction seems to have little effect on the diffusion range. An inspection of the 
forming adlayer however reveals that its morphology depends on how strongly adatoms interact 
with each other relative to the quasicrystaline substrate atoms. As the relative interaction 
strength of the adatoms is increased three dimensional cluster-type growth and island formation 
is observed. As we showed before, the structure of the adlayer either is an imperfect 
continuation of the quasicrystalline substrate maintaining local five fold coordination 
(QC-like phase), or it consists of nano-scale fcc crystallites with their (111) surface parallel 
the decagonal surface of the quasicrystaline substrate, and whose azimuthal orientations 
in equal increments of $36^\circ$ results in a pseudodecagonal symmetry (FCC domain phase) 
\cite{mungan,sven}. Our simulations show that the QC-like phase emerges at fast thermal 
relaxation times and weak adatom interactions, with the growth mode being layer by-layer. 
The FCC domain phase occurs when either the relaxation times or the relative strength 
of the adatom interactions is increased along with a transition from layer by layer growth to 
a cluster type growth that seems to depend predominantly on the thermal relaxation times. 

We are thus able to reproduce numerically the formation and size selection of domains, their distinct orientations and have shown how the depositional dynamics of adatoms on the substrate determines 
its morphology. Our experimental and computational efforts in determining the structure of growing adlayers have shown that there is a structural registry between the crystal and the quasicrystal on a local scale.  Experimentally observed diffraction patterns contain an orientational smearing of the structure with a well-defined length scale associated with adatom spacing and symmetry. 

The results indicate that epitaxial oxygen atoms chemisorbed on the quasicrystalline surface experience competing interactions. While they favor ordering in a stable hexagonal mesh, the quasicrystalline surface template they condense on, force them into an aperiodic order. The system finds the best compromise by partially satisfying both conditions and breaks up into domains ordered in a hexagonal structure, where each domain remains locally commensurate and in registry with the substrate. This registry results in five distinct orientations for the oxygen nanocrystals. The size of these domains is determined by the interfacial strain energy, as this size increases the island edges get rapidly out of registry with the substrate and it becomes energetically more favorable to break-up into locally commensurate domains.

In heteroepitaxial systems the interfacial strain energy increases with the film thickness, and beyond a critical thickness strain built in the film causes the film to relax to its stable bulk phase by creating misfit dislocations at the interface. In monoatomic systems on decagonal quasicrystalline surfaces, this critical thickness does not seem to extend beyond a monolayer thickness, and the system relaxes into large hexagonal domains. It is likely, however that, in some carefully chosen binary and ternary alloy systems the critical thickness may increase significantly, and one may be able to grow epitaxy stabilized quasicrystalline phases, that do not exist in nature.

Computations were mainly done using the  Gilgamesh and Kassandra computer clusters at the Feza G\"ursey Institute and Bo\u gazi\c ci University. This work has been funded in part by grant 08B302 of Bo\u gazi\c ci University. Financial support by Schweizerischer Nationalfonds is gratefully appreciated.

\end{document}